\documentclass[aps,superscriptaddress,nofootinbib,eqsecnum,prd,notitlepage,twocolumn, showkeys]{revtex4-1} 

\pdfoutput=1

\usepackage{amsfonts}
\usepackage{amsmath}
\usepackage{amssymb}
\usepackage{graphicx,color}
\usepackage{float}
\usepackage{hyperref}
\usepackage{subfigure}
\usepackage{dcolumn}
\usepackage{soul}
\usepackage{ulem}
\usepackage{verbatim}

\begin{document}

\title{Stability analysis of warm quintessential dark energy model}

\author{Suratna Das}
\email{suratna.das@ashoka.edu.in}
\affiliation{Department of Physics, Ashoka University,
   Rajiv Gandhi Education City, Rai, Sonipat: 131029, Haryana, India}

\author{Saddam Hussain}
\email{msaddam@iitk.ac.in}
\affiliation{Department of Physics, Indian Institute of Technology, Kanpur, Uttar Pradesh 208016, India}

\author{Debottam Nandi}
\email{dnandi@physics.du.ac.in}
\affiliation{Department of Physics and Astrophysics, University of Delhi, Delhi 110007, India}

\author{Rudnei O. Ramos} \email{rudnei@uerj.br}
\affiliation{Departamento de Fisica Teorica, Universidade do Estado do
  Rio de Janeiro, 20550-013 Rio de Janeiro, RJ, Brazil }

\author{Renato Silva}
\email{renatobqa@gmail.com}
\affiliation{Departamento de Fisica Teorica, Universidade do Estado do
  Rio de Janeiro, 20550-013 Rio de Janeiro, RJ, Brazil }

\begin{abstract}

A dynamical system analysis is performed for a model of dissipative
quintessential inflation  realizing warm inflation at early primordial
times and dissipative interations in the dark sector at late
times. The construction makes use of a generalized exponential
potential realizing both phases of accelerated expansion. A focus is
given on the behavior of the dynamical system at late times and the
analysis is exemplified by both analytical and numerical results.  The
results obtained demonstrate the viability of the model as a
quintessential inflation model in which stable solutions can be
obtained.

\end{abstract}

\maketitle

\section{Introduction}

Cosmic inflation~\cite{Kazanas:1980tx,Guth:1980zm,Sato:1981ds,Sato:1980yn,Linde:1981mu,Albrecht:1982wi}, 
proposed as a
solution to the fine-tuning problems of the Big Bang theory, describes
an accelerated expanding phase in the early Universe. It is also
commonly assumed to be driven by a potential energy dominated scalar
field, called the inflaton. On the other hand, the observational
discovered late-time cosmic acceleration of our
Universe~\cite{SupernovaCosmologyProject:1998vns,SupernovaSearchTeam:1998fmf}
can also be explained by the dynamics of a scalar field, called the
quintessence \cite{Peebles:1987ek,Ratra:1987rm} (for reviews, see, e.g., 
 \cite{Bamba:2012cp,Tsujikawa:2013fta}). There has been constant effort
in the literature to unify the early- and the late-time cosmic
accelerations by making the same scalar field play the role of both
the inflaton and the quintessence field (for recent reviews, see
\cite{deHaro:2021swo,Bettoni:2021qfs} and references therein). However, the main
obstacle to unify the early- and the late-time accelerations by a single
scalar field is that, as in conventional cold inflation, the energy
density in the inflaton field must, at least partially, decay to 
radiation at the end of inflation in order to reheat the
Universe, while part of the energy density of the inflaton must survive until 
recently if the inflaton should also play the role of quintessence. 
To overcome this difficulty, a number of alternative reheating
mechanisms have been proposed, such as gravitational
reheating~\cite{Ford:1986sy,Chun:2009yu}, instant
preheating~\cite{Felder:1998vq,Campos:2002yk}, curvaton
reheating~\cite{Feng:2002nb,BuenoSanchez:2007jxm},
non-minimal~\cite{Dimopoulos:2018wfg} or Ricci
reheating~\cite{Bettoni:2018utf, Opferkuch:2019zbd}, just to cite some examples.   

Another novel way of overcoming the reheating problem in such unified
models is to opt for warm inflation (WI)~\cite{Berera:1995ie}  as the
inflationary model (for recent reviews on WI, see, e.g.,
\cite{Kamali:2023lzq,Berera:2023liv}). WI is a variant
inflationary scenario where the inflaton field, having strong
couplings with other fields, dissipates its energy to a thermal bath
during inflation. As a constant thermal bath is maintained throughout
WI, it smoothly ends in a radiation dominated Universe, without
invoking the need of a separate reheating phase. Thus, WI can naturally
alleviate the problem of reheating in such unified quintessential
inflation models. Besides, the constraints set by the swampland
conjectures (especially the de Sitter
conjecture~\cite{Ooguri:2018wrx,Garg:2018reu}), which prohibit
constructions of de Sitter vacua in string theory, cannot be easily
met by the conventional inflationary
models~\cite{Agrawal:2018own, Kinney:2018nny}. WI, however, naturally overcome those
constraints~\cite{Das:2018hqy,Motaharfar:2018zyb,Das:2018rpg,Berera:2023mlj}
and, thus, can be considered as a viable inflationary paradigm in
string landscapes. On the late-time acceleration front, quintessence
is in better agreement with the swampland conjectures than a non-zero
cosmological constant $\Lambda$~\cite{Agrawal:2018own}.\footnote{It is to note that, in general, quintessence dark energy models, preferred by the swampland conjectures~\cite{Agrawal:2018own}, exacerbates the $H_0$ tension as have been pointed out in~\cite{Colgain:2019joh,Banerjee:2020xcn}. {}For some of the recent discussions concerning the differences on the value of $H_0$
between the measurements coming from the Cosmic Microwave Background~\cite{Planck:2018vyg} and by local distance
measurements~\cite{Riess:2016jrr,Riess:2018byc,Riess:2019cxk}, see, e.g., Refs.~\cite{Dominguez:2019jqc, Park:2019emi, Lin:2019zdn, Freedman:2020dne, Birrer:2020tax, Boruah:2020fhl, Freedman:2021ahq, Wu:2021jyk, Cao:2022ugh}.} Hence,
unifying WI with quintessence has an added advantage even from the
point of view of effective field theories consistent with a quantum
gravity ultraviolet realization. 

The first two attempts~\cite{Dimopoulos:2019gpz,Rosa:2019jci}
made in the literature to unify WI with late-time
quintessence-driven acceleration,
dissipative effects played a role only during the early-time
inflationary phase, after which they die down when WI
ends. Afterwards, it was assumed that the late-time acceleration was 
driven by standard
quintessence dynamics, where the quintessence field is treated to be
decoupled from the rest of the matter in the Universe. Above all, two
different forms of potentials of the same scalar field are required to
drive the two accelerating  phases, at early- and at late-times, in these
models. There has been another novel attempt to unify WI and the
late-time acceleration~\cite{Lima:2019yyv}, where the scalar field
first dissipates its energy to a radiation bath during inflation and,
at a later stage, due to couplings with matter, it dissipates its energy to
the matter content of the Universe. This non-relativistic matter
content, generated due to the dissipation of the scalar field, is
shown to be able to account for the dark matter in the Universe. Besides, in
the implementation considered in Ref.~\cite{Lima:2019yyv} only one
form of the scalar potential (a generalized form of exponential potential) is
required to drive both the early- and the late-time accelerations,
which is an added advantage. Thus, this model accounts for inflation,
dark matter and dark energy at one go\footnote{A double-field warm inflation model was also recently been proposed \cite{DAgostino:2021vvv}, where inflation, dark matter and dark energy can be realised in a single setup.}.

The main feature of WI, which distinguishes it from the standard
inflationary paradigm, is the dissipative effects of the inflaton
field during inflation. The presence of dissipation makes WI a rich
dynamical system, whose stability in the early Universe has been previously
analyzed in the literature~\cite{deOliveira:1997jt,Moss:2008yb,delCampo:2010by,Bastero-Gil:2012vuu, Li:2018sfs}.
However, a study of how a similar analysis could be carried out when those dissipative
effects can extend up to the late-times, is still largely missing. In
the unified model described in \cite{Lima:2019yyv}, dissipation
effects are effective even after the inflationary phase, and
interactions in the post-inflationary epoch are motivated fully from
the WI picture. Though this might have similarities with models describing
interactions in the dark sector (see, e.g., the review papers
\cite{Bolotin:2013jpa,Wang:2016lxa}), the model studied here is,
however, much more reminiscent of the WI idea, but extending it to quintessential
inflation models. Therefore, we will call the late-time acceleration of
this model as {\it warm quintessential dark energy model}. The aim of this
paper is to perform the first study in the literature of the stability of the 
dynamical system of this warm quintessential dark energy model. 

We have organized this paper as follows. In Sec.~\ref{model}, we
discuss the model whose stability we want to determine in this
paper. In Sec.~\ref{dyn-sys}, we discuss the dynamical system produced 
by the model and show that it indeed accounts for four different
phases of evolution: (a) inflation, (b) radiation domination, (c)
matter domination and (d) dark energy domination for some generic
choices of parameters. In Sec.~\ref{qualitative}, we qualitatively show
that the late-time acceleration is an attractor solution of the
model. In the following section, Sec.~\ref{quantitative}, we perform a
rigorous dynamical system analysis to  study the stability of the
system depending on different model parameters.  In
Sec.~\ref{conclusion}, we discuss our main results and
conclude. {}Finally, an Appendix is included where we also study the
stability of the slow-roll trajectories in both early- and late-time
epochs.

\section{Model} 
\label{model}

In our model, the quintessential scalar field decays to both radiation
and matter energy densities.  Here, we propose the complete set of
background equations involving the quintessential scalar field
$\phi$, the radiation fluid energy density $\rho_r$ and the matter energy
density $\rho_m$, with evolution equations as given, respectively, by
\begin{eqnarray}
&& \ddot \phi + 3H \dot \phi +\Upsilon_r \dot \phi + \Upsilon_m \dot
  \phi + V_{,\phi}=0,
\label{eqphinew}
\\ && \dot \rho_r + 4 H \rho_r = \Upsilon_r  \dot \phi^2 ,
\label{eqrhoRnew}
\\ && \dot \rho_m + 3 H \rho_m = \Upsilon_m \dot \phi^2,
\label{eqrhom}
\end{eqnarray}
where $V_{,\phi}$ is the field derivative of the quintessential scalar
potential, $\Upsilon_r$ describes the energy exchange  between the
quintessential scalar field and radiation energy density, 
$\Upsilon_m$ describes the energy exchange  between the quintessential
scalar field and matter energy density and the Hubble parameter is given by
the Friedmann equation,
\begin{equation}
H^2\equiv\frac{\dot a^2}{a^2}= \frac{1}{3 M_{\rm Pl}^2} \left( \frac{\dot \phi^2}{2} + V
+ \rho_r + \rho_m\right), 
\label{Hub}
\end{equation}
with $a$ the scale factor and $M_{\rm Pl} \equiv (8\pi G)^{-\frac{1}{2}} \simeq 2.44\times
10^{18}$ GeV  is the reduced Planck mass and $G$ is Newton's
gravitational constant. 

We parametrize the
dissipation terms $\Upsilon_r$ and $\Upsilon_m$ in the following generic forms, 
which are motivated from many early works on WI and 
also discussed in Ref.~\cite{Lima:2019yyv},

\begin{eqnarray}
\Upsilon_r = C_r \rho_R^{c/4} \phi^p M^{1-c-p},
\label{Upsilonr}
\end{eqnarray}
and
\begin{eqnarray}
\Upsilon_m = C_m \rho_m^{k/4} \phi^q M^{1-k-q},
\label{Upsilonm}
\end{eqnarray}
where $C_r$ and $C_m$ are dimensionless constants and $M$ is some
appropriate (constant) scale with mass dimension. Hence,
$[\Upsilon_r]=M$ and $[\Upsilon_m]=M$ \footnote{Note  that in
principle we do not need to have both dissipation terms with the same
mass scale and we could define them with different scales.  But any
difference between these scales can be absorbed in the dimensionless
constants $C_r$ and $C_m$ anyway.}. The various powers $c,\;
p,\;k,\;q$ model the different dependencies that these dissipation
coefficients might have with the quintessence background field,
radiation energy density and
matter energy density. These parameters are not all arbitrary and the
stability of the dynamical system can put strong bounds on them, as we will
see. In particular, the stability of the system under
slow-roll demands that $|c|<4$ and $|k|<4$
(see the Appendix~\ref{stability-slow-roll} for details). 

Appropriate choices of
dependencies on  $\phi$, $\rho_r$ and $\rho_m$ can be made in
Eqs.~(\ref{Upsilonr}) and (\ref{Upsilonm}) such that we can have, for
example, the dissipation coefficient $\Upsilon_r$, given in
Eq.~(\ref{Upsilonr}), dominating during inflation, thus leading to a
WI regime, while $\Upsilon_m$, given in  Eq.~(\ref{Upsilonm}), only
dominates at late times~\cite{Lima:2019yyv}. While $\Upsilon_m$ can be  subdominant at
primordial times, it can help in setting an initial abundance for the
matter density.  Given appropriate parameters $C_m$ and $M$, we can
arrange for the matter-quintessence scalar field to display a similar
behaviour found, e.g., in the case of nonminimal couplings of the
scalar field  to matter~\cite{Amendola:1999er}, thus modelling
different energy exchange forms between the dark sector components.
The matter-quintessence scalar field interaction term, under
appropriate choices of parameters, can also help in providing an extra
friction force on the quintessence scalar field and, thus, help making
$\phi$ acquire a negative equation of state at late times, signalling
the beginning of the dark energy (quintessence) domination epoch
and even making scalar fields with steeper potentials more likely to work 
as a quintessence field, as we will see later.

\section{The dynamical system} 
\label{dyn-sys}

The evolution equations~(\ref{eqphinew}) - (\ref{eqrhom}) can be
brought into a form appropriate for a dynamical system analysis by
defining the variables~\cite{Bahamonde:2017ize}
\begin{eqnarray}
&& x= \frac{\dot \phi}{\sqrt{6} M_{\rm Pl} H},
\label{defx}
\\ && y= \sqrt{\frac{V}{3}}\frac{1}{ M_{\rm Pl} H},
\label{defy}
\\ && \Omega_r= \frac{\rho_r}{3 M_{\rm Pl}^2 H^2},
\label{defr}
\\ && \Omega_m= \frac{\rho_m}{3 M_{\rm Pl}^2 H^2}.
\label{defm}
\end{eqnarray}
Note that from the above definitions, we have
that
\begin{equation}
x^2+y^2 = \Omega_\phi,
\label{Omegaphi}
\end{equation}
is the fraction in energy density of the quintessence scalar field.
{}From Eqs.~(\ref{defx}) - (\ref{defm}), the {}Friedmann equation~(\ref{Hub})
becomes equivalent to
\begin{equation}
1= x^2 + y^2 + \Omega_r +\Omega_m.
\label{constraint}
\end{equation}

The evolution equations (\ref{eqphinew}) - (\ref{eqrhom}) can then be
brought into a dynamical system form as
\begin{eqnarray}
x'&=&-\frac{3 x (1-x^2)}{2}+\frac{\Omega_r x}{2}-3x ( Q_m + Q_r)
\nonumber \\ &+&\sqrt{\frac{3}{2}} \lambda y^2-\frac{3 x y^2}{2},
\label{eqx}
\\ y'&=&\frac{3 y}{2}+\frac{3 x^2 y}{2}-\frac{3
  y^3}{2}-\sqrt{\frac{3}{2}} x y \lambda +\frac{y \Omega_r}{2},
\label{eqy}
\\ \lambda'&=&-\sqrt{6} x (-1+\Gamma ) \lambda^2,
\label{eqlambda}
\\ \Omega_r'&=& 6 x^2 Q_r-\Omega _r+3 x^2 \Omega _r-3 y^2 \Omega
_r+\Omega _r^2,
\label{eqOmegar}
\end{eqnarray}
where, in the above equations, a prime means derivative with respect to
the number of efolds, $'\equiv d/dN$, where $dN= H dt$, while $Q_r$
and $Q_m$ are the dissipation ratios, defined as
\begin{equation}
Q_r =\frac{\Upsilon_r}{3H},
\label{Qr}
\end{equation}
and
\begin{equation}
Q_m=\frac{\Upsilon_m}{3H}.
\label{Qm}
\end{equation}
In Eqs.~(\ref{eqx}) - (\ref{eqOmegar}) we have also introduced the
variable $\lambda$, which is defined as
\begin{equation}
\lambda = -M_{\rm Pl} \frac{V_{,\phi}(\phi)}{V(\phi)},
\label{lambda0}
\end{equation}
and $\Gamma$ in Eq.~(\ref{eqlambda}) is defined as
\begin{equation}
\Gamma= \frac{V(\phi) V_{,\phi\phi}(\phi)}{V_{,\phi}^2(\phi)}.
\label{Gamma}
\end{equation}

Note that the Eqs.~(\ref{eqx}) - (\ref{eqOmegar}) are general and
valid in principle  for any potential. To complete the dynamical
system, we also need the evolution equations for the dissipation
ratios $Q_m$ and $Q_r$ and to fix the form of the inflaton potential
$V(\phi)$. {}For definiteness, let us consider the generalized
exponential inflaton potential of the form:
\begin{equation}
V(\phi) = V_0 e^{-\alpha (\phi/M_{\rm Pl})^n},
\label{pot}
\end{equation}
where $V_0$ is the normalization of the potential, $\alpha$
is a dimensionless constant here taken as positive and $n>1$
for potentials steeper than the simple exponential potential.
This form of potential was originally proposed in Ref.~\cite{Geng:2015fla}
and considered also in Refs.~\cite{Geng:2017mic,Ahmad:2017itq,Shahalam:2017rit,Das:2019ixt}
for quintessential inflation in the cases of absence of dissipation
(i.e., radiation production). The first use of this potential 
in the context of warm quintessential inflation was in 
Ref.~\cite{Lima:2019yyv} and later also considered in 
Refs.~\cite{Gangopadhyay:2020bxn,Basak:2021cgk}. Studies involving 
observational predictions for this model in the context of WI
were developed in Refs.~\cite{Das:2020xmh,Das:2022ubr}. 

{}From the potential (\ref{pot}), we then obtain that 
\begin{equation}
\Gamma=1-\frac{(n-1)}{n \alpha } \left(\frac{\lambda }{n \alpha
}\right)^{\frac{n}{1-n}},
\end{equation}
and
\begin{equation}
\lambda= n \alpha  \left(\frac{\phi }{M_{\text{Pl}}}\right)^{n-1}.
\label{lambda}
\end{equation}
Note that for $n\neq 1$, $\phi$ is related to $\lambda$ by
\begin{equation}
\phi=M_{\rm Pl} \left(\frac{\lambda}{n \alpha}\right)^{\frac{1}{n-1}}.
\label{phi}
\end{equation}
{}From the above definitions, the evolution equations for $Q_m$ and $Q_r$ can be expressed,
respectively, as
\begin{eqnarray}
Q_m'&=& \frac{3 (2-k)Q_m}{4} +\frac{3 (x^2-y^2) Q_m}{2} \nonumber
\\ &+&\sqrt{6} q x  \left(\frac{n \alpha }{\lambda
}\right)^{\frac{1}{-1+n}} Q_m+\frac{Q_m \Omega _r}{2} \nonumber \\ &-&
\frac{3 k x^2 Q_m^2}{2 \left(-1+x^2+y^2+\Omega _r \right)},
\label{eqQm}
\\ Q_r' &=& \frac{3 (1-2c)Q_r}{2} +\frac{3 (x^2 -y^2)Q_r}{2} \nonumber
\\ &+& \sqrt{6}  p x \left(\frac{n \alpha }{\lambda
}\right)^{\frac{1}{-1+n}} Q_r+ \frac{3 c x^2 Q_r^2}{2 \Omega _r}
\nonumber \\ &+&\frac{Q_r \Omega _r}{2}.
\label{eqQr}
\end{eqnarray}

In writing the system of equations Eqs.~(\ref{eqx}) -
(\ref{eqOmegar}), (\ref{eqQm}) and (\ref{eqQr}),  we have considered
the fraction in energy density in matter as equivalently to the first
integral  of Eq.~(\ref{eqrhom}) and which is determined through the
constraint Eq.~(\ref{constraint}).  The system of equations
Eqs.~(\ref{eqx}) - (\ref{eqOmegar}), (\ref{eqQm}) and (\ref{eqQr}),
together with Eq.~(\ref{constraint}), hence, form a complete set of
equations describing the dynamics of the system.

\begin{center}
\begin{figure}[!htb]
\includegraphics[width=7.5cm]{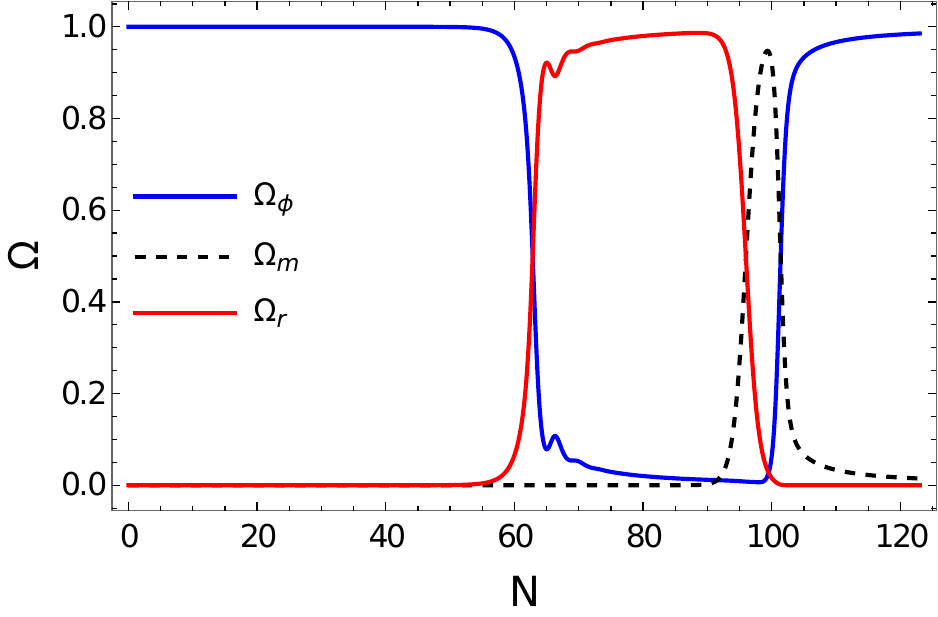}
\caption{The energy density ratios as a function of the number of
  e-folds fos an  inflaton potential with constants $n=3$ and
  $\alpha=0.015$. The initial conditions considered were such that
  $x(0) = 0.0025,\, \Omega_m(0)=10^{-50},\, \Omega_r(0)= 5.8 \times
  10^{-10}, \, \lambda(0) = 6.1 \times 10^{-3}$, while for the
  dissipation coefficient ratios we have  considered $Q_r(0) =
  10^{-4}$, $Q_{m,1}(0) = 1.2 \times 10^{-40}$ for the ratio
  corresponding to  the dissipation coefficient $\Upsilon_{m,1}$ and
  $Q_{m,2}(0) = 1.7 \times 10^{-65}$ for the dissipation coefficient
  $\Upsilon_{m,2}$ (the combination of the two  dissipation
  coefficients were considered such to reproduce the analogous case of
  Ref.~\cite{Lima:2019yyv}).}
\label{fig1}
\end{figure}
\end{center}

In Ref.~\cite{Lima:2019yyv}, the evolution equations (\ref{eqphinew})
- (\ref{eqrhom}) were solved assuming a dissipation coefficient
$\Upsilon_r$ given by
\begin{equation}
\Upsilon_r=  C_r \rho_r^{3/4}/\phi^2,
\label{Upsr}
\end{equation}
while $\Upsilon_m$ was taken to be of the form $\Upsilon_m= \Upsilon_{m,1}
+\Upsilon_{m,2}$, where
\begin{eqnarray}
\Upsilon_{m,1}&=&C_m \rho_m^{3/4}/\phi^2,
\label{Upsm1}
\\ \Upsilon_{m,2} &=& M^2/\rho_m^{1/4}.
\label{Upsm2}
\end{eqnarray}
Let us show that in this case, the dynamical system given by
Eqs.~(\ref{eqx}) - (\ref{eqOmegar}), (\ref{eqQm}) and (\ref{eqQr})
lead to the same dynamics as shown in Ref.~\cite{Lima:2019yyv}. In
{}Fig.~\ref{fig1} we show the  result obtained by the solution of the
dynamical system for the energy density fractions $\Omega_\phi,\,
\Omega_r$ and $\Omega_m$ and which is obtained by a representative
example of initial conditions. We see that the system of equations
(\ref{eqx}) - (\ref{eqOmegar}), (\ref{eqQm}) and (\ref{eqQr}) produce
an evolution that is initially  characterized by an accelerated
inflationary regime, when $\Omega_\phi$ dominates. This phase smoothly
goes to a radiation dominated regime when $\Omega_r$
dominates. Towards the end of the evolution, it displays a short
matter dominated phase, before $\Omega_\phi$ becomes the dominating
component again in the future, which corresponds to a dark energy
phase.

\begin{center}
\begin{figure}[!htb]
\includegraphics[width=7.5cm]{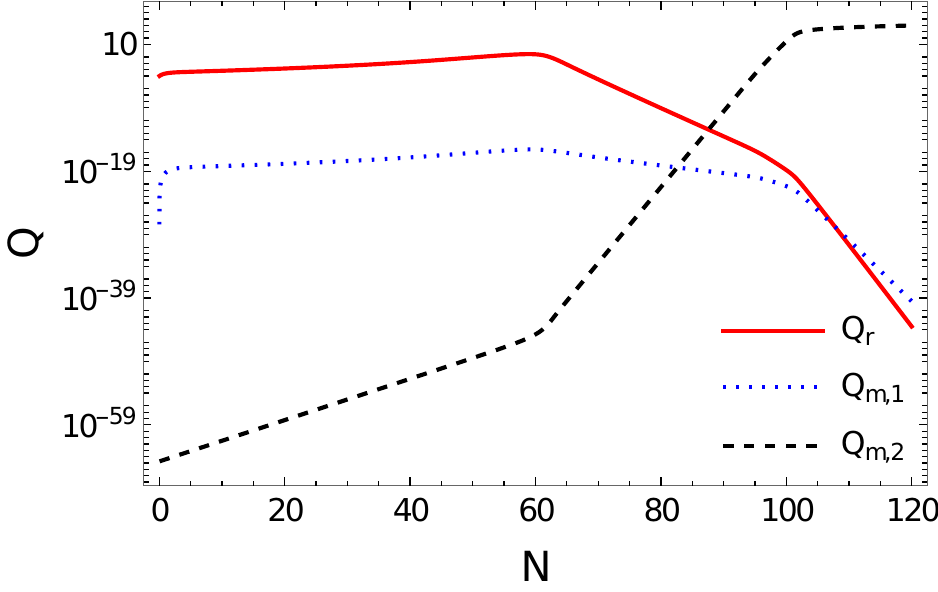}
\caption{The evolution of the dissipation ratios $Q_r$, $Q_{m,1}$ and
  $Q_{m,2}$.}
\label{fig2}
\end{figure}
\end{center}

In {}Fig.~\ref{fig2}, we give the evolution of the dissipation ratios
$Q_r$, $Q_{m,1}$ and $Q_{m,2}$ obtained from Eqs.~(\ref{Upsr}),
(\ref{Upsm1}) and (\ref{Upsm2}). Note that right after inflation both
$Q_r$ and $Q_{m,1}$ drop similarly, while $Q_{m,2}$ is enhanced after
inflation, during the radiation era, while flattening at late
times. This shows that different choices of the powers in
Eqs.~(\ref{Upsilonr}) and (\ref{Upsilonm}), can lead to different
evolutions during different epochs in the universe.

\section{Late time dynamics: A qualitative analysis}
\label{qualitative}

Analyzing the complete dynamical system made of the Eqs.~(\ref{eqx}) -
(\ref{eqOmegar}), (\ref{eqQm}) and (\ref{eqQr}) is too complicated
given that it is a six-dimensional order system. However, we can still
get valuable information looking at snapshots of the system on a given
plane. The most interesting plane to look at is the plane $(x,y)$,
which gives us information about the behavior of the trajectories
passing through the accelerated region.  This is of particular
importance when studying the late-time dynamics of the system, where
we want to know about the ability of the system in reaching a DE
dominated regime.  We perform this analysis next, and leave the full
dynamical system analysis of the late-time dynamics for the next
section. 

Since we are interested in the late-time behavior of the system, we
can ignore the radiation related terms in Eqs.~(\ref{eqx}) and
(\ref{eqy}), which can then be approximated as
\begin{eqnarray}
x'&\simeq&-\frac{3 x (1-x^2)}{2}-3x Q_m  \nonumber
\\ &+&\sqrt{\frac{3}{2}} \lambda y^2-\frac{3 x y^2}{2},
\label{eqxlate}
\\ y'&\simeq&\frac{3 y}{2}+\frac{3 x^2 y}{2}-\frac{3
  y^3}{2}-\sqrt{\frac{3}{2}} x y \lambda ,
\label{eqylate}
\end{eqnarray}
with the constraint that
\begin{equation}
x^2 + y^2 \leq 1,
\label{lateconstr}
\end{equation}
and the trajectories in the plane $(x,y)$ are then constrained to be
in the semi-circle defined by Eq.~(\ref{lateconstr}) and $y\geq 0$
(meaning positive potential energy).  At fixed values of $\lambda$ and
$Q_m$, the fixed points derived from Eqs.~(\ref{eqxlate}) and
(\ref{eqylate}) are given by
\begin{eqnarray}
&&P_1 = (0,0),
\label{P1}
\\ &&P_2=(-\sqrt{1+2Q_m},0),
\label{P2}
\\ &&P_3=(\sqrt{1+2Q_m},0),
\label{P3}
\\ &&P_4=(x_4,y_4),
\label{P4}
\end{eqnarray}
where
\begin{eqnarray}
x_4&=& \frac{3+\lambda ^2+3 Q_m}{2 \sqrt{6} \lambda } \nonumber \\ &-&
\frac{\sqrt{\lambda ^4+6 \lambda ^2 \left(-1+Q_m \right)+ 9
    \left(1+Q_m \right){}^2}}{2 \sqrt{6} \lambda },
\label{x4}
\\ y_4 &=&  \left\{\frac{6 \lambda ^2-\lambda ^4+9 \left(1+Q_m
  \right){}^2}{12 \lambda^2} \right.  \nonumber \\ &+&
\left. \frac{\left[\lambda ^2-3 \left(1+Q_m \right)\right] }{12
  \lambda^2} \right.  \nonumber \\ &\times& \left. \sqrt{\lambda^4+6
  \lambda ^2  \left(-1+Q_m \right)+9 \left(1+Q_m
  \right){}^2}\right\}^{\frac{1}{2}}.  \nonumber \\
\label{y4}
\end{eqnarray}
It can be checked that both points $P_2$ and $P_3$ are repelling
nodes, while $P_1$ is a saddle. The point $P_4$ is an attractor. 

It is
useful to look at the asymptotic  value for $P_4$ for large
$\phi$. {}From Eq.~(\ref{lambda}) we then have that $\lambda \to 0$.
Expanding the point $P_4$ for $\lambda \ll 1$, we obtain that its
coordinates in the plane $(x,y)$ satisfy
\begin{equation}
x_4 \sim \frac{\lambda}{\sqrt{6} (1+Q_m)} + {\cal O}(\lambda^3),
\end{equation}
and
\begin{equation}
y_4 \sim 1- \frac{(1+2 Q_m) \lambda^2}{12(1+Q_m)^2} + {\cal
  O}(\lambda^4).
\end{equation}
Thus, the point $P_4$ is an attractor for the trajectories leading,
asymptotically, to a dark energy accelerating regime, with the
potential energy of the quintessence field dominating at later times. 

\begin{center}
\begin{figure}[!htb]
\includegraphics[width=7.5cm]{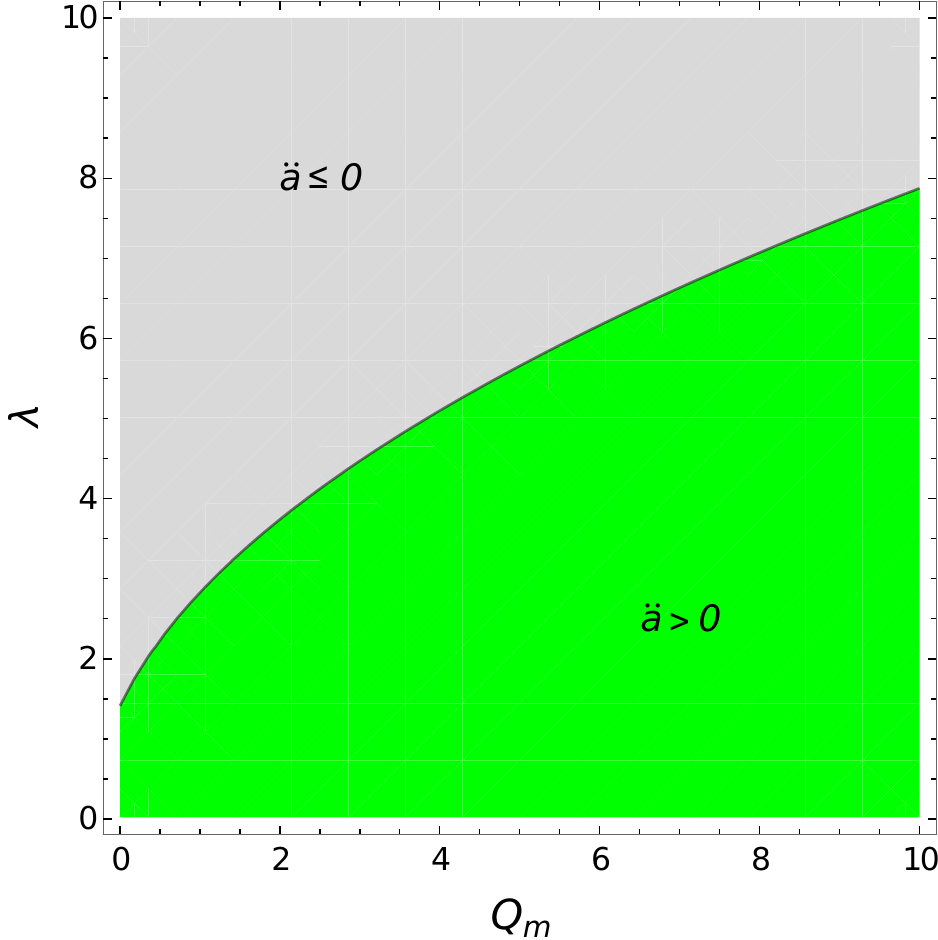}
\caption{The region of parameters $\lambda$ and $Q_m$ allowing for
  acceleration at late times.}
\label{fig3}
\end{figure}
\end{center}

Note that the larger is  $Q_m$, the easiest is to enter in the
accelerating regime, with $x_4\to 0$ and $y_4\to 1$ for $Q_m \gg 1$
and $\lambda \ll 1$. Physically, the dissipation term $\Upsilon_m$
acts as a friction term slowing down the quintessence field at later
times and making it easier to enter the accelerating regime $\ddot a
>0$.

\begin{center}
\begin{figure*}[!htb]
\subfigure[]{\includegraphics[width=7.5cm]{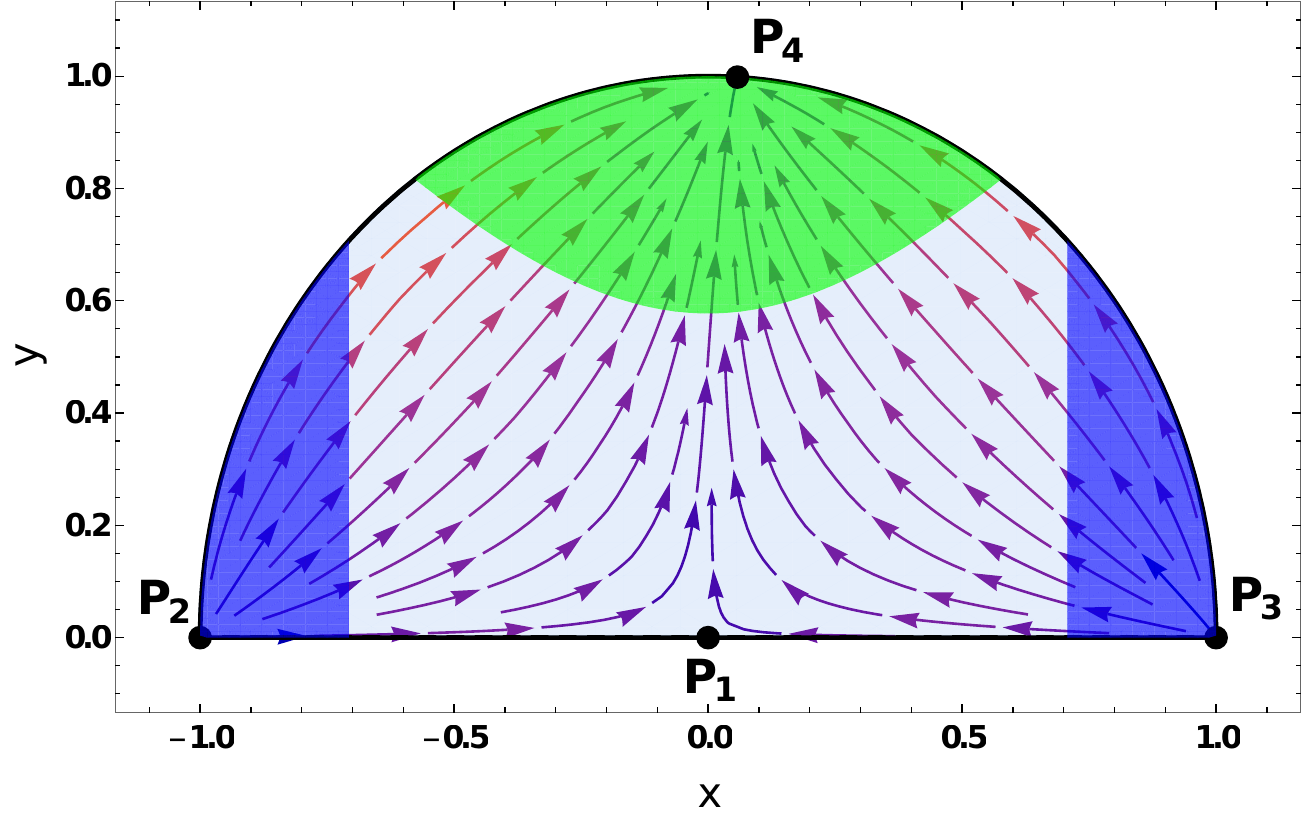}}
\subfigure[]{\includegraphics[width=7.5cm]{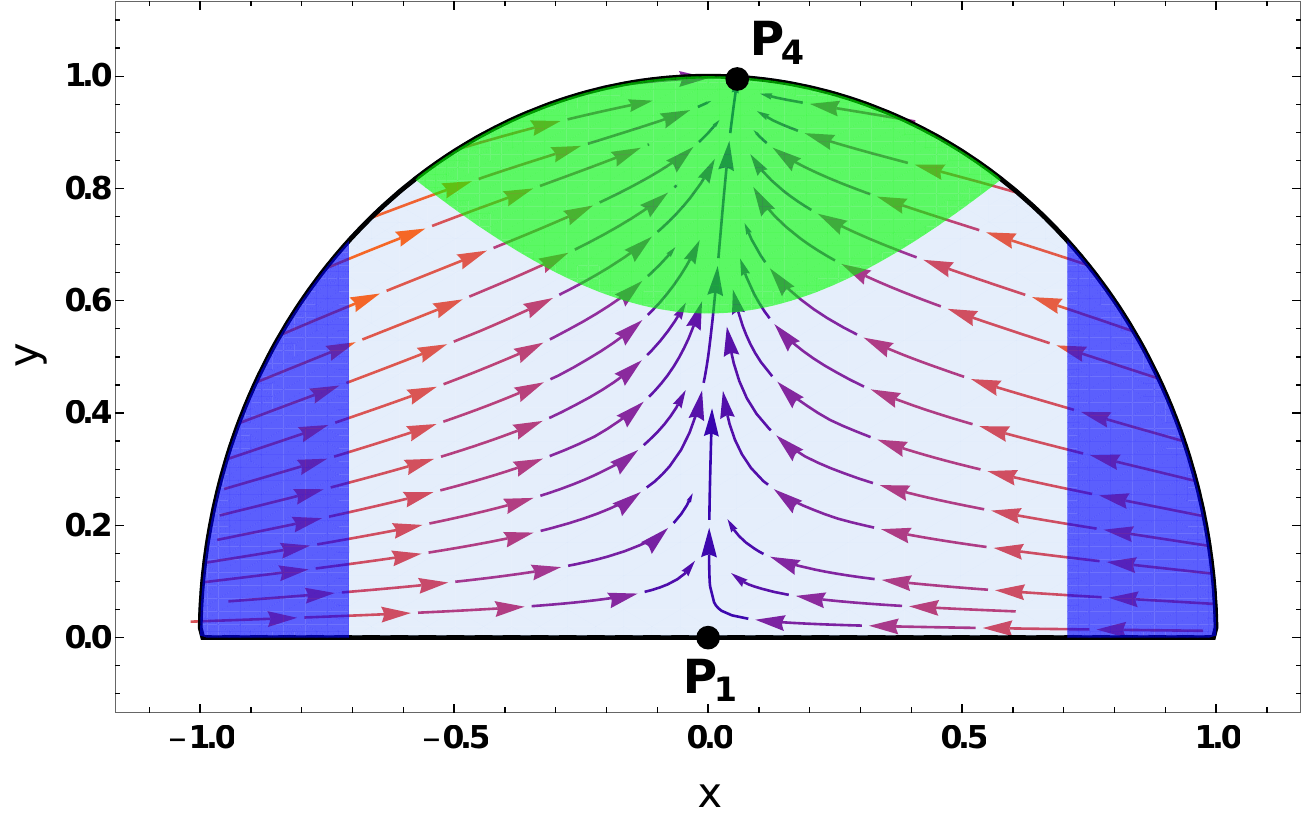}}
\subfigure[]{\includegraphics[width=7.5cm]{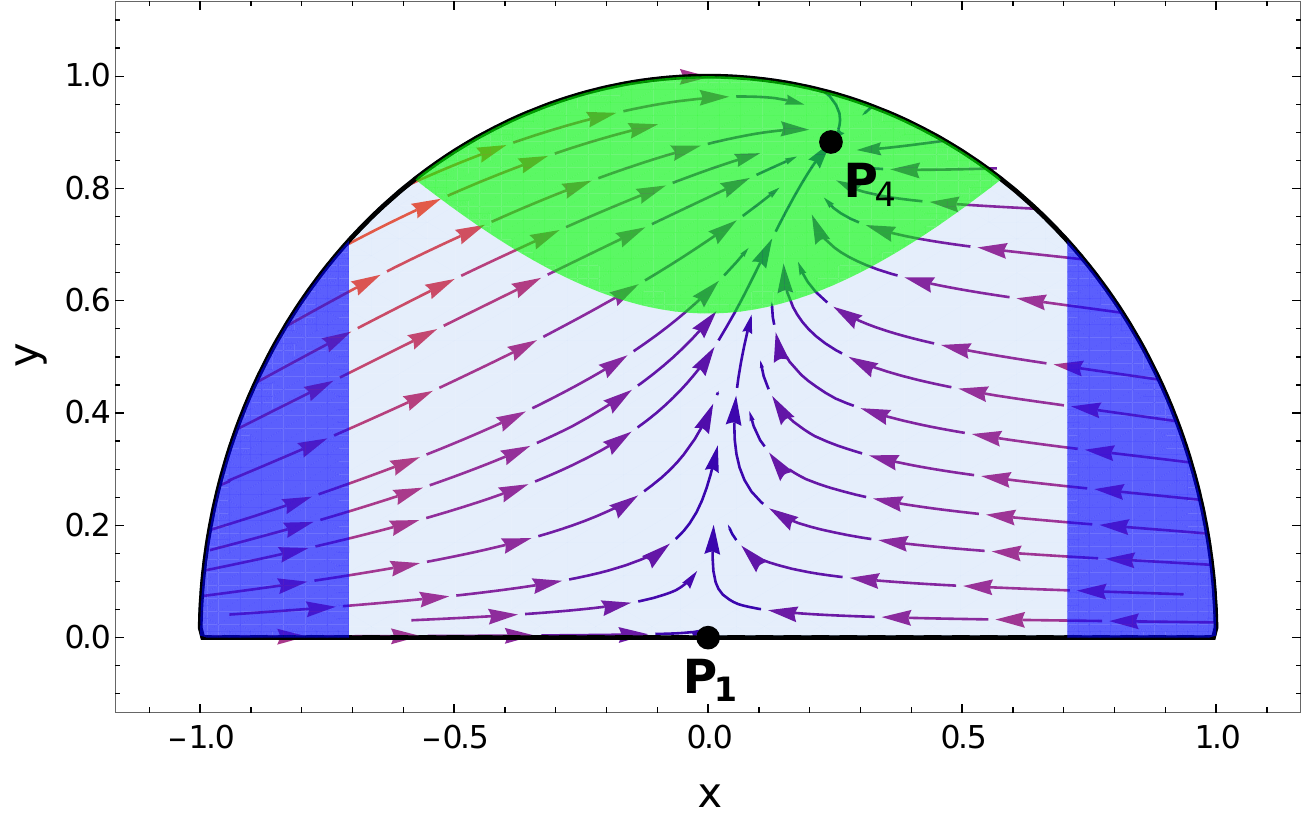}}
\subfigure[]{\includegraphics[width=7.5cm]{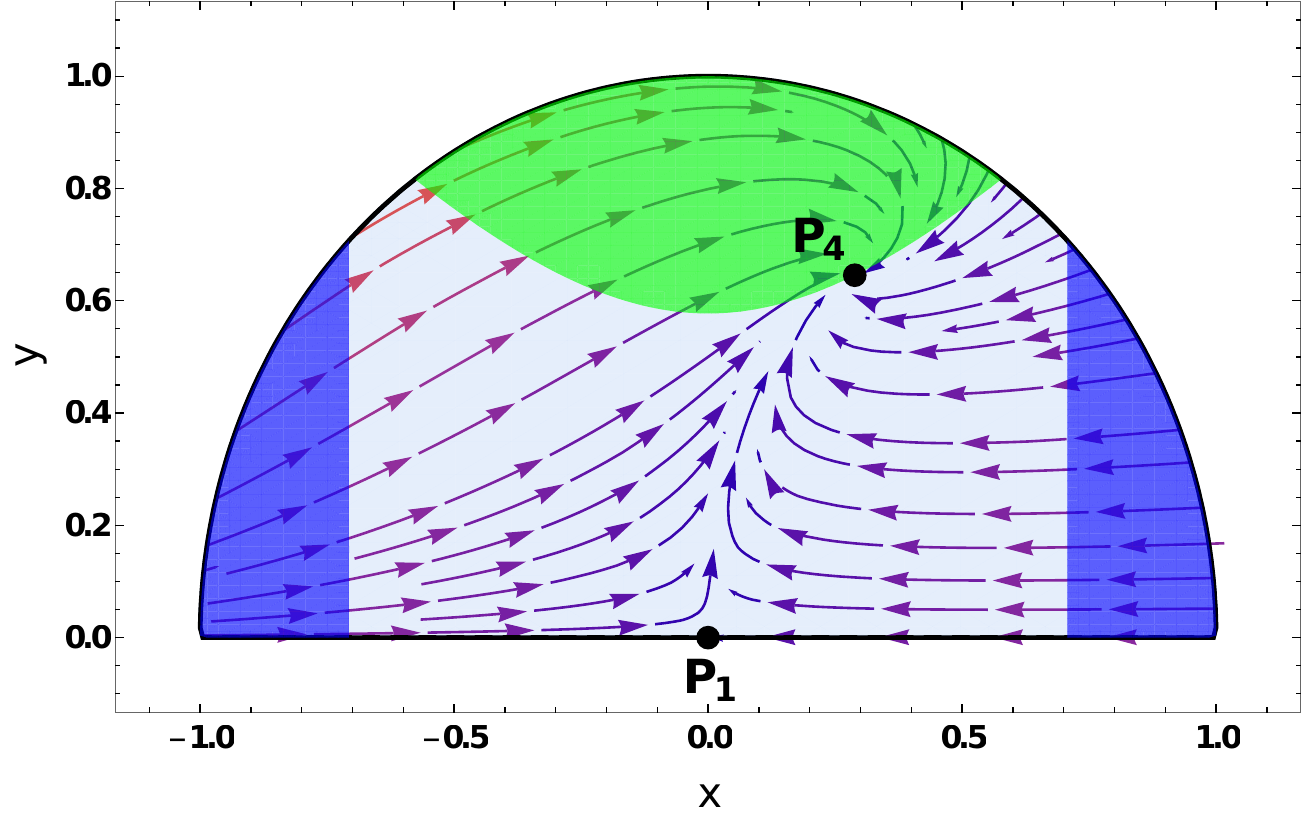}}
\subfigure[]{\includegraphics[width=7.5cm]{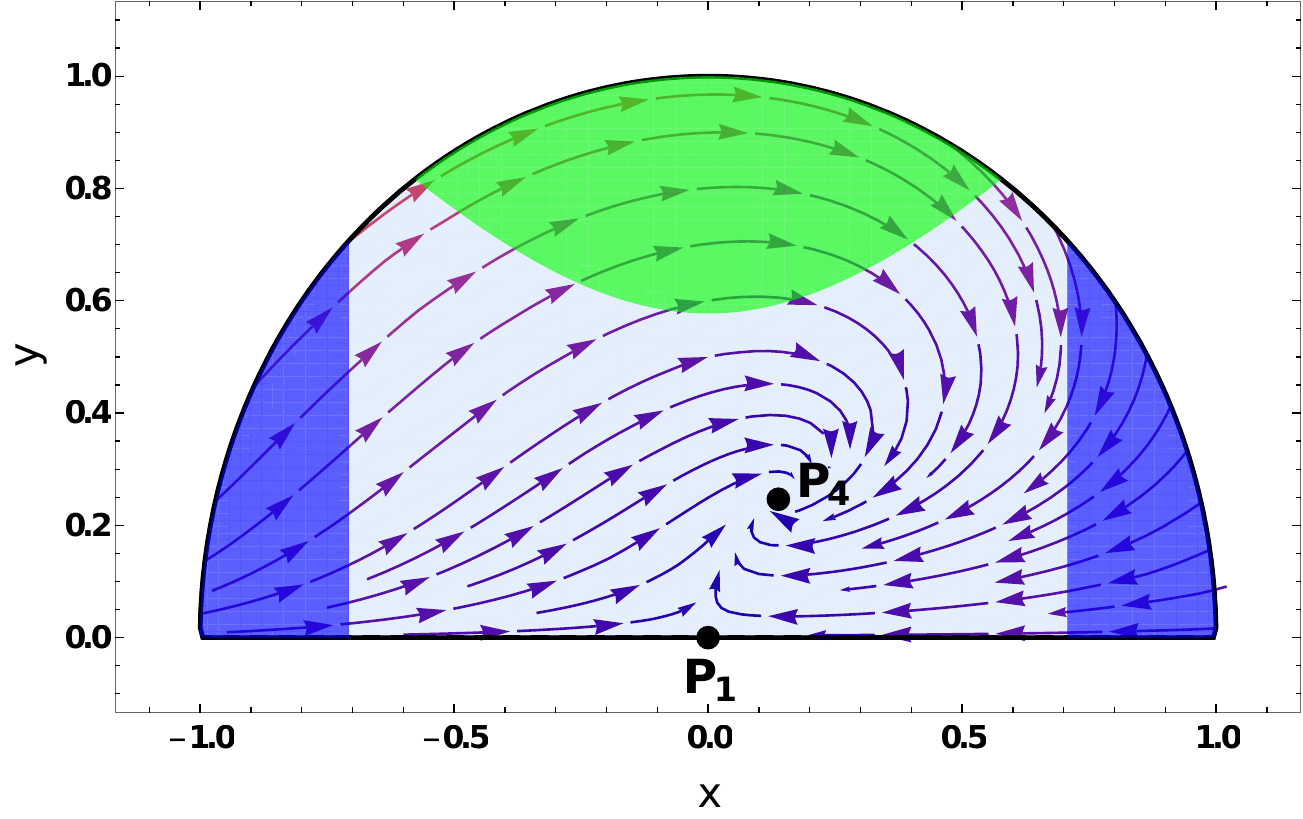}}
\caption{Snapshots of the phase space trajectories of the dynamical
  system in the plane $(x,y)$ for different values of parameters.
  Panel a: $Q_m=0$, $\lambda=\lambda_{\rm accel}/10$; Panel b:
  $Q_m=1$, $\lambda=\lambda_{\rm accel}/10$; Panel c:  $Q_m=1$,
  $\lambda=\lambda_{\rm accel}/2$; Panel d:  $Q_m=1$,
  $\lambda=\lambda_{\rm accel}$; Panel e:  $Q_m=1$, $\lambda=3
  \lambda_{\rm accel}$.}
\label{fig4}
\end{figure*}
\end{center}

The region in the parameters $Q_m$ and $\lambda$ for which the point
$P_4$ is in the accelerating regime is illustrated in
{}Fig.~\ref{fig3}. Note that the larger values of $Q_m$ allow for 
steeper potentials to work as quintessence fields (i.e., allowing for
later accelerated regimes). 
The boundary of the accelerating and nonaccelerating regions shown in
{}Fig.~\ref{fig3} is given by the condition $x_4^2-y_4^2 = -1/3$, {\it
  i.e.}, where the equation of state is exactly $\omega=-1/3$. It is
found to be given by
\begin{equation}
\lambda_{\rm accel} =\sqrt{2+6Q_m}.
\label{lambdaaccel}
\end{equation}

In {}Fig.~\ref{fig4}, we give snapshots of the phase space
trajectories of the dynamical system in the plane $(x,y)$ for
different values of parameters.  The green shaded region denotes the
accelerating region, the blue shaded region is the kination region,
where the kinetic energy of the quintessence field dominates, and
which satisfies $x^2>1/2$. The gray region is the matter dominated
region.

Note that for $Q_m>0$, the points $P_2$ and $P_3$ move away from the
boundary of the region shown in {}Fig.~\ref{fig4}. This is why they
are  not shown in {}Fig.~\ref{fig4} panels (b)-(e).  Since $P_2$ and
$P_3$ lie in the kination regions, the trajectories then emanate from
the blue region. Note also that as $\lambda$ increases (for a fixed
value of $Q_m$), the point $P_4$ moves from the accelerating region
and towards the point $P_1$ for matter domination.  It is interesting
to look at the corresponding value of $\lambda_{\rm accel}$ from the
example given in {}Fig.~\ref{fig1}. At later times, the total
dissipation ratio $Q_m=Q_{m,1}+Q_{m,2}$ flattens with a value
$Q_m\simeq 10^4$. {}From  Eq.~(\ref{lambdaaccel}), the corresponding
value for $\lambda_{\rm accel}$ is then $\lambda_{\rm accel} \sim
240$. The value of $\lambda$ at later times in the case of the initial
conditions used in the example of {}Fig.~\ref{fig1} is $\lambda_f \sim
22 \ll \lambda_{\rm accel}$.  Thus, the system at later times goes to
the accelerating dark energy dominated regime as expected.

\section{Late time dynamics: full dynamical system analysis}
\label{quantitative}

The quintessential scalar field predominantly dissipates to matter
energy density during the late times as it is evident from
{}Fig.~\ref{fig1}. Therefore, for the late-time dynamics, one can
ignore the contribution of the radiation bath ($\Omega_r\approx0$),
which means that we can also ignore the equations of $\Omega_r$ and
$Q_r$ as given in Eqs.~(\ref{eqOmegar}) and (\ref{eqQr}),
respectively. Hence, the previous dynamical system of six equations
now reduces to a dynamical system of four equations:
\begin{eqnarray}
x'&=&-\frac{3 x (1-x^2)}{2}-3x Q_m +\sqrt{\frac{3}{2}} \lambda
y^2-\frac{3 x y^2}{2},
\\ y'&=&\frac{3 y}{2}+\frac{3 x^2 y}{2}-\frac{3
  y^3}{2}-\sqrt{\frac{3}{2}} x y \lambda ,
\\ Q_m'&=& \frac{3 (2-k)Q_m}{4} +\frac{3 (x^2-y^2) Q_m}{2}
\nonumber\\ 
&\!\!\!\!+& \!\!\!\!\! \sqrt{6} q x  \left(\frac{n \alpha }{\lambda
}\right)^{\frac{1}{-1+n}} Q_m- \frac{3 k x^2 Q_m^2}{2 \left(-1+x^2+y^2
  \right)},
\\ \lambda'&=&\sqrt{6}x(n-1)(\alpha
n)^{\frac{1}{n-1}}\lambda^{\frac{n-2}{n-1}},
\end{eqnarray}
and the constraint equation given in Eq.~(\ref{constraint}) becomes
\begin{eqnarray}
1=x^2+y^2+\Omega_m.
\label{constraint1}
\end{eqnarray}
We define the equation of state of the total fluid (including both
$\Omega_m$ and $\Omega_\phi$) as 
\begin{eqnarray}
\omega_{\rm tot}\equiv \frac{p_\phi+p_m}{\rho_\phi+\rho_m}=x^2-y^2.
\end{eqnarray}

Before analyzing the set of autonomous equations, we note that, though
$x$ and $y$ are bounded between -1 to 1 by the constraint given in
Eq.~(\ref{constraint1}), $\lambda$ and $Q_m$ are unbounded and can
take values between 0 to $\infty$. To obtain dynamical parameters
which are bounded, unlike $\lambda$ and $Q_m$, it becomes convenient
to introduce two new variables $z$ and $\xi$ that are defined as
\begin{eqnarray}\label{z_def}
	z &=& \dfrac{\lambda^\frac{1}{n-1}}{1+
          \lambda^\frac{1}{n-1}},  \\ \xi &=& \frac{Q_m}{1+ Q_m},
\end{eqnarray}
which make $z$ and $\xi$ range from  $0 < z <1$ and $ 0< \xi < 1$.
However, we found that the transformed dynamical set of equations in
terms of $(x, y, z, \xi)$ displays only the trivial critical point
$(0,0,z,0)$, while the other possible critical points, 
including any accelerating solutions, remain hidden. 
This seems to be a drawback of the choice of variables made, but that 
can be overcame as follows. To work around the above mentioned difficulty,
 we first redefine $z$ as 
\begin{equation}\label{z_new_def}
	z= \dfrac{1-\lambda^\frac{1}{n-1}}{ \lambda^\frac{1}{n-1}},  
\end{equation}
which now ranges from $-1<z\leq0$ for values $1\leq\lambda<\infty$. Here
we note that for values of $\lambda$ smaller than unity, $z$ again
becomes unbounded which we do not want. That is why we restrict the
lower value of $\lambda$ to 1. 

Next, we can make a nontrivial transformation of the variables $z$ and
$\xi$ to two other parameters $u$ and $v$ as
\begin{equation}\label{u-and-v}
	u  =  \frac{\xi - z}{\xi + z}, \quad v  = \frac{\xi -
          z}{\xi^2}\,, 
\end{equation}
such that 
\begin{eqnarray}
\label{}
	\xi = \frac{2u}{(1 + u)v } , \quad z = \frac{ 2 u (1-u)}{(1 +
          u)^2 v}.
\end{eqnarray}
The $\xi$ and $z$ variables ranges (which are, respectively, given by 
$0<\xi<1$ and $-1<z\leq 0$) put
constraints on the $u$ and $v$ values as 
\begin{eqnarray}\label{uv_const}
&& u \le -1 \implies v > \frac{2 (u-1) u}{(u+1)^2}, \nonumber \\ &&u
  \ge 1 \implies v>\frac{2 u}{u+1}.
\end{eqnarray}
Therefore, in terms of the four variables $(x,y,u,v)$, we get the set
of autonomous equations as follows,
\begin{widetext}
\begin{eqnarray}\label{}
	x' &=& \frac{3 y^2}{\sqrt{6}} \left( \frac{v (1+u)^2}{v(1+u)^2
          + 2 u(1-u)}\right)^{n-1}+ \dfrac{3}{2} x\left(-2 -  \dfrac{4
          x  u}{v(1+u)-2u} +   1+x^2-y^2
        \right),\label{new_x_prime}\\ 
        y' & =& \dfrac{- x y  \sqrt{6}
        }{2}\left( \frac{v(1+u)^2}{v(1+u)^2 + 2 u(1-u)}\right)^{n-1} +
        \dfrac{3y}{2} \left( 1 + x^2 -y^2\right), \label{new_y_prime}\\
         u' &=& \frac{1}{4} \bigg[\frac{3 k x^2 }{2 \Omega_m }
          \frac{4 u^2 (1-u)}{v} + \frac{2u (1-u)}{v} ((1+u) v-2u  )
          \bigg\{\frac{-3 k }{4} + \frac{q \sqrt{6} x ( \alpha
            \ n)^{\frac{1}{n-1}} }{(1+u)^2 v} \left((1+u)^2 v + 2u - 2
          u^2\right) \nonumber \\   &&+ \frac{3}{2} \left( 1 + x^2
          -y^2\right)	\bigg\}  + \frac{\sqrt{6}x \left((1 + u)^2 v
            + 2 u (1-u)\right)^2 ( \alpha \ n)^{\frac{1}{n-1}} }{(1+
            u)v} \bigg], \label{u_prime}\\ 
            v' & =& \frac{3 k x^2 }{2
          \Omega_m }u + \frac{(v(1+u) -2u)}{2} \left[\dfrac{-3 k}{4} +
          q \sqrt{6} x (\alpha \  n)^{\frac{1}{n-1}} \frac{\left((1 +
            u)^2 v  + 2 u -2u^2  \right)}{(1+u)^2 v} + \frac{3}{2}
          \left( 1 + x^2 -y^2\right)\right] \nonumber\\ & &+\sqrt{6} x
        (\alpha \  n)^{\frac{1}{n-1}} \frac{\left((1+u)^2 v + 2 u
          (1-u)\right)^2}{4u (1+u)^2} -\frac{6 k x^2 }{ \Omega_m }
        \dfrac{u}{(1+u)} - \frac{2((1+u)v - 2u)}{(1+u)} \bigg[
          -\frac{3 k}{4} + q \sqrt{6} x (\alpha n)^{\frac{1}{n-1}}
          \nonumber \\ &&\times  \dfrac{\left((1+u)^2 v + 2u -
            2u^2\right)}{(1+u)^2 v} + \frac{3}{2} \left( 1 + x^2
          -y^2\right) \bigg]\,, \label{v_prime} 
\end{eqnarray}
\end{widetext}
where $\Omega_m=1-x^2-y^2$ from Eq.~(\ref{constraint1}). It is to note
that the dynamical system analysis with steeper exponential potentials
($n>1$) is a tasking job, as has been pointed out in \cite{Das:2019ixt}. The linear stability
analysis \cite{Bahamonde:2017ize} breaks down in such cases as the
real parts of some of the eigenvalues turn out to be zero. In
\cite{Das:2019ixt}, the authors used the center manifold theorem
to analyze the stability of a system with steep exponential
potentials. However, for the present problem we found that with dissipation of the scalar
field to the matter sector, the system becomes too intricate to be
analyzed employing the center manifold theorem technique. Therefore,
we chose the non-trivial parametrization, given in
Eq.~(\ref{u-and-v}), which enables us to do the stability analysis of
the system with dissipation.

The non-trivial transformation in Eq.~(\ref{u-and-v}) has created a
complicated structure of the  autonomous system, making it difficult
to identify the critical points analytically. Therefore, we shall
compute the fixed points by assuming some representative values of the
model parameters \( (k, q , n, \alpha)\). The advantage of this
non-trivial transformation is that we can find non-trivial fixed
points and the linearization technique~\cite{Bahamonde:2017ize} does
not break down, allowing us to find non-zero eigenvalues.  We select
the fixed points based on the constraints $ 0 \le x^2 + y^2 \le 1$
and $ 0 < \xi < 1,\  -1 < z  \le 0$. 

Before we proceed to find the fixed points and their corresponding
eigenvalues, we note that in Eqs.~\eqref{u_prime} and
\eqref{v_prime}, the term \( x^2/\Omega_{m}\) is discontinuous at
the point $(x=0, y=1)$,
\begin{equation}\label{}
	\lim_{\substack{x\to 0 \\ y\to 1}} \frac{x^2}{1-x^2-y^2} = 0,
        \quad  	\lim_{\substack{y\to 1 \\ x\to 0}}
        \frac{x^2}{1-x^2-y^2} = -1.
\end{equation} 
However, this discontinuity can be removed by multiplying the above
expression by \(x\).  Therefore, in the autonomous equations this can
be achieved by redefining the time variable \( {dN}\to x\  {dN}\). This
redefinition does not change the dynamics of the system and it removes
the discontinuity.  Therefore, the structure of the new autonomous system
becomes 
\begin{equation}\label{}
	\frac{d \vec{x}}{dN} = f(\vec{x}) \times x,
\end{equation}
where \(\vec{x} = \{x,y, u,v\}\). This set of autonomous equations is
now suitable for finding completely all the critical points. The critical points for
four different example cases have been evaluated in
Table~\ref{tab:redefined_critical_pts}. To determine the critical
points and for illustration, we have fixed the parameters $n$ and
$\alpha$ of the potential as $n=3$ and $\alpha=0.015$ as have been
considered in Ref.~\cite{Lima:2019yyv}, while four different
representative choices for $k$ and $q$ are made.  The motivation for the
choice of parameters come from the fact, as shown in Refs.~\cite{Lima:2019yyv} and \cite{Das:2020xmh}, 
that the type of runaway exponential potential that we have studied here satisfies well
the observations (e.g., the tensor-to-scalar ratio $r$ and spectral tilt $n_s$)
for the cases for which the power $n$ in the potential is equal to or
larger than 2. Hence, we have fixed in our examples the case $n=3$ as a representative
case, ensuring that the warm inflationary dynamics can correctly
satisfy the Planck results for $r$ and $n_s$. The same reason motivated our 
different choices for the constant $\alpha$ in the potential, while the
choices for $k$ and $q$, that controls the dissipation in the dark energy regime,
were chosen in analogy to the similar powers ($c$ and $p$) appearing in the dissipation
coefficient during the inflationary regime.  We discuss
the stability of these four chosen cases below.

\begin{table}[t]
	\centering
	\begin{tabular}{|c|c|c|c|c|c|}
		\hline
		\hline
		Points & $ (x,y,u,v) $ & $ \omega_{\rm tot} $ & $\Omega_{\phi}$ & $\Omega_{m}$ & Stability\\
		\hline
		\hline
		\multicolumn{6}{c}{\textbf{Case I:}\ $\bf  n =3,\ k = 3, \ q = -2,\ \alpha = 0.015 $  }\\
		\hline
		\hline
		$ M_{0} $ & $ (0,\ any, \ any, \ any)  $ & $-y^2$ & $y^2$& $1-y^2$ & Stable\\
		\hline 
		$M_{1}$ & $(-0.06,0.96,1.44,1.19)$ & $-0.92$ & $0.92$ & $0.08$ & Saddle\\
		\hline
		$M_{2}$ & $(0.34,0.82,2.02,2.69)$ & $-0.56$ & $0.79$ & $0.21$& Saddle\\
		\hline
		$M_{3}$ & $(0.17,0.23, 7.11,2.01)$ & $-0.06$ & $0.12$ & $0.88$ & Saddle\\
		\hline	
		$M_{4}$ & $(-0.16,0.04,3.59,1.93)$ & $0.02$ & $0.03$ & $0.97$ & Saddle \\
		\hline		
		\hline
		\multicolumn{6}{c}{\textbf{Case II:}\ $ \bf n =3,\ k = -1, \ q = 0,\ \alpha = 0.015$}\\
		\hline
		\hline
		$M_{0}$ & $(0,\ any,\ any,\ any)$ & $-y^2$ & $y^2$ & $1-y^2$ & Stable\\
		\hline 			
		\multicolumn{6}{c}{\textbf{Case III:}\ $ \bf n =3,\ k = 0, \ q = 0, \alpha = 0.001$}\\
		\hline
		\hline
	$M_{0}$ & $(0.20,0.84,1.99,1.45)$ & $-0.66$ & $0.74$ & $0.26$ & Stable\\
	\hline
	$M_{1}$ & $(0.39,0.87,1.99,4.33)$ & $-0.60$ & $0.91$ & $0.09$ & Saddle\\
		\hline 	
	\multicolumn{6}{c}{\textbf{Case IV:}\ $ \bf n =3,\ k = 3, \ q = 0, \alpha = 0.015$}\\
	\hline
	\hline
	$M_{0}$ & $(0,any,any,any)$ & $-y^2$ & $y^2$&$1-y^2$ & Stable \\
			\hline
		$M_{1}$ & $(0.38,0.86, 1.99,3.32)$ & $-0.59$& $0.88$ & $0.12$ & Saddle\\
		\hline
		$M_{2}$ & $(0.09,0,1.99,1.66)$ & $0.01$ & $0.01$ & $0.99$ & Saddle\\
		\hline
		\hline
	\end{tabular}
	\caption{Critical points of the redefined autonomous system. }
	\label{tab:redefined_critical_pts}
\end{table}

\subsection{Case I: $\ n =3,\ k = 3, \ q = -2,\ \alpha = 0.015 $}

{}Firstly, we consider the model with $k=3$ and
$q=-2$, which corresponds to the dissipation coefficient
$\Upsilon_{m,1}$ (and therefore $Q_{m,1}$) given in
Eq.~(\ref{Upsm1}). According to {}Fig.~\ref{fig2}, this dissipation
coefficient is responsible for the decay of the quintessence field into
matter during the early phases of the evolution. In this case, we
found the five critical points as given in Table~\ref{tab:redefined_critical_pts}.  

\begin{center}
 \begin{figure}[!htb]
 	\centering \includegraphics[width=10cm]{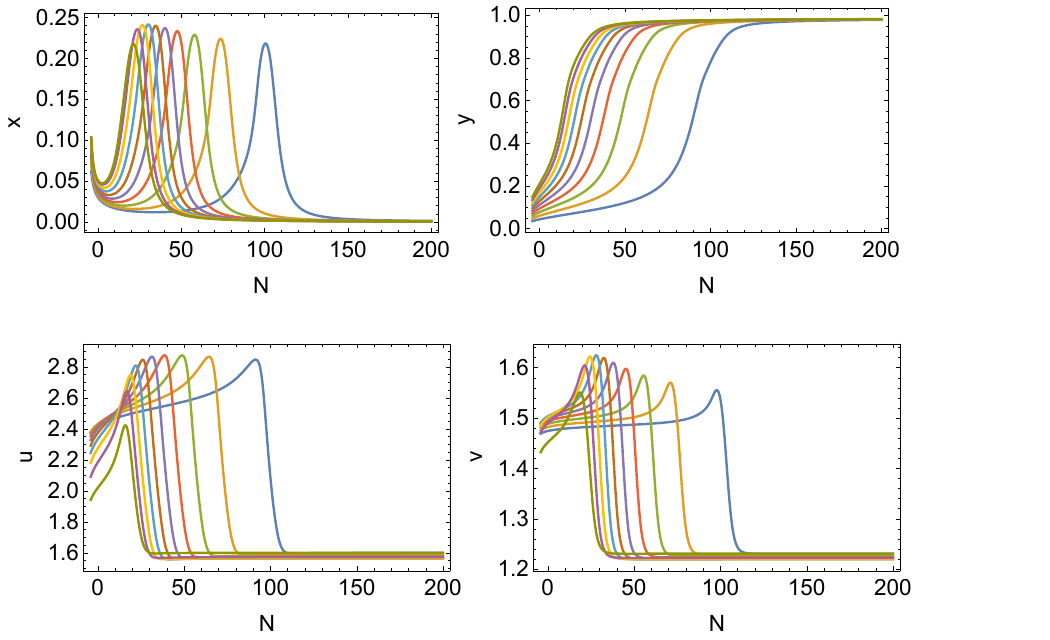}
 	\caption{Numerical evolution of the autonomous equations for
          the critical point $M_0$  with \(k=3, q=-2, n=3, \alpha =
          0.015\). The dynamical parameters $x$, $y$, $u$, and $v$ 
have been evolved numerically with ten different initial conditions. 
The ten different colored lines in each of these panels represent the 
evolution of these parameters with these varied initial conditions. 
The initial values of each of these parameters can be read from each 
of the plots at $N=0$.  }
 	\label{fig:numeric_evo_k3_alp015}
 \end{figure}
\end{center}

\begin{figure}[!htb]
	\centering \includegraphics[width=7.5cm]{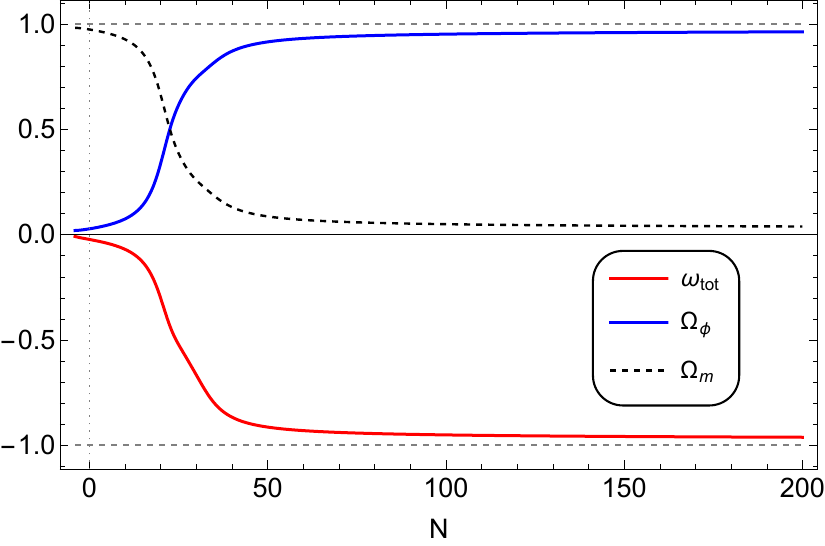}
	\caption{Evolution of cosmological parameters \( (\omega_{\rm
            tot},\ \Omega_{\phi}, \  \Omega_{m})\) corresponds to the
          critical point $M_0$ with $  k =3, \ q = -2, \ n =3,\ \alpha
          = 0.015$.  }
	\label{fig:evo_k3_a015}
\end{figure}

The critical points $M_{1}$
  and $M_2$ indicate accelerating solutions with equation of state
  given by $\omega_{\rm tot}=-0.92$ and $\omega_{\rm tot}=-0.56$,
  respectively. In these cases, the scalar field density dominates
  over the matter energy density. However, we found both these points
  to be saddles.  The critical points $M_3$ and $M_4$ indicate matter
  domination, $\omega_{\rm tot}\sim0$, with matter density dominating
  over the scalar field density. Both of these points turn out to be
  saddles too.  At the fifth critical point, \( M_{0}\), the eigenvalues turn out 
  to be zero, the conventional
  linearization technique is no longer applicable. Hence, the
  stability for this critical point has to be determined numerically
  by varying the initial conditions. 
  If $x=0$ initially, \((y, u,v)\)
  can take any value maintaining the constraint relations $ 0 \le
  x^2 + y^2 \le 1$, $0 < \xi < 1$ and  $-1 < z  \le 0$. 
We then evolve the system numerically
  and the evolution of the dynamical parameters $(x,y,u,v)$ is depicted in
  the {}Fig.~\ref{fig:numeric_evo_k3_alp015}. We show in this  figure that, 
even if we choose the initial values away from the critical point, 
they converge to $(x,y,u,v)\simeq (0, 1, 1.6, 1.2)$ as time goes by, 
ensuring a stable solution at  late times. We plot the evolution of
  the cosmological parameters $\omega_{\rm tot}$, $\Omega_\phi$ and
  $\Omega_m$ for this critical point in
  {}Fig.~\ref{fig:evo_k3_a015}. It is seen that $\omega_{\rm tot}$ tends 
to $-1$ steadily at later times and the energy density is fully dominated by 
the scalar field density $\Omega_\phi$. {}Figures~\ref{fig:numeric_evo_k3_alp015} 
and \ref{fig:evo_k3_a015} confirm that the
  critical point $M_0$ is stable and yields an accelerating solution
  with $\omega_{\rm tot}\sim-1$. 

\subsection{Case II: $ n =3,\ k = -1, \ q = 0,\ \alpha = 0.015$}

\begin{center}
 \begin{figure}[!htb]
 	\includegraphics[width=10cm]{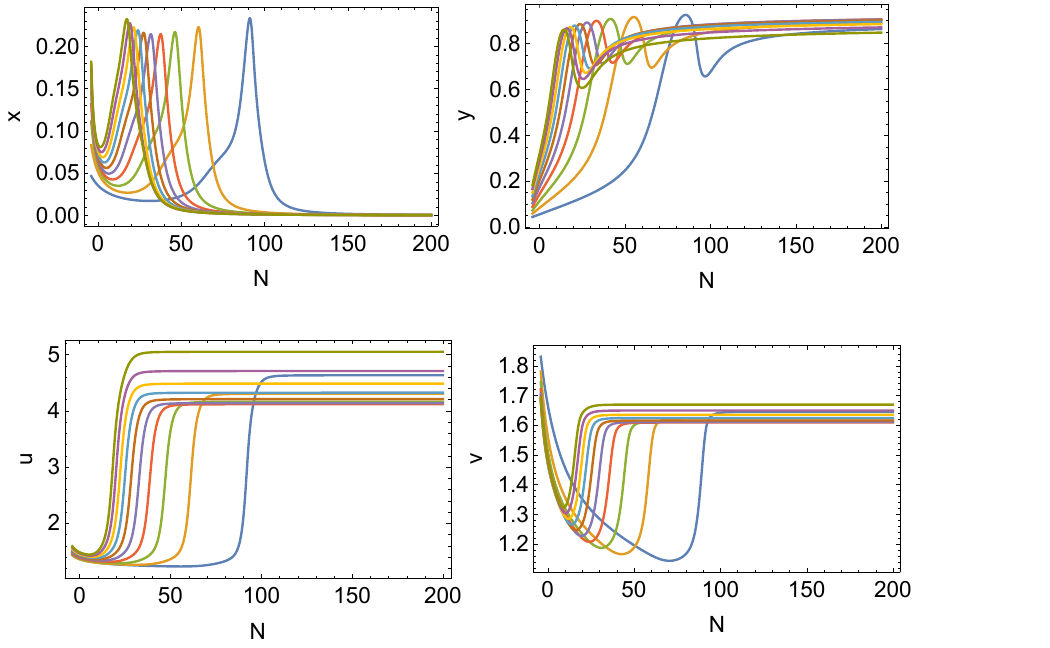}
 	\caption{Numerical evolution of the autonomous equations for
          the critical point $M_0$  with \(k=-1, q=0, n=3, \alpha =
          0.015\).The dynamical parameters $x$, $y$, $u$, and $v$ 
have been evolved numerically with ten different initial conditions. 
The ten different colored lines in each of these panels represent the 
evolution of these parameters with these varied initial conditions. 
The initial values of each of these parameters can be read from each 
of the plots at $N=0$.  }
 	\label{fig:numeric_evo_km1_a015}
 	 \end{figure}
	 \end{center}

\begin{figure}[!htb]
	\centering \includegraphics[width=7.5cm]{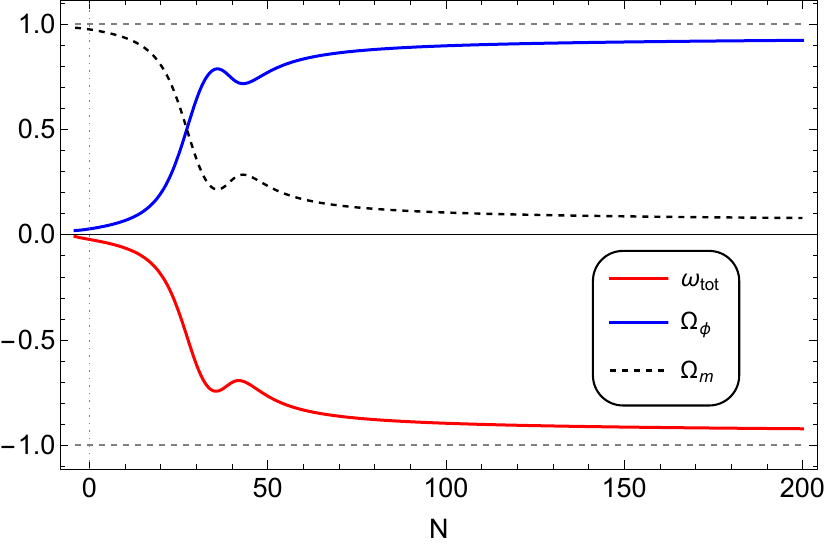}
	\caption{Evolution of cosmological parameters \( (\omega_{\rm
            tot},\ \Omega_{\phi}, \  \Omega_{m})\) corresponds to the
          critical point $M_0$ with $ k =-1, \ q = 0, \ n =3,\ \alpha
          = 0.015$.  }
	\label{fig:evo_km1_new}
\end{figure}

Here, we consider for illustration the model with
  $k=-1$ and $q=0$, which correspond to the dissipation coefficient
  $\Upsilon_{m,2}$ (and therefore $Q_{m,2}$) given in
  Eq.~(\ref{Upsm2}). According to {}Fig.~\ref{fig2}, this dissipation
  coefficient is responsible for the decay of the quintessence field into
  matter during late times. We found only one critical point $M_0$ for
  this case. As before, the eigenvalues for this critical point turn
  out to be zero. Hence, we then resort again to a numerical analysis of 
the stability for this point. Initially,  \((y, u,v)\) can take any value
  maintaining the constraint relations $0 \le x^2 + y^2 \le 1$,
$0 < \xi < 1$ and $-1 < z  \le 0$, with fixed $x=0$. We evolve the system
  numerically and the corresponding evolution of the dynamical parameters 
$(x,y,u,v)$ is depicted in {}Fig.~\ref{fig:numeric_evo_km1_a015}. 
We see that although $x, y$ and $v$ converge at late times, $u$ does not 
converge to a single value. Still, the system does not diverge and, thus, 
shows stability at late times. To establish the stability of this point, we 
further plot the evolution of the cosmological parameters $\omega_{\rm tot}$,
  $\Omega_\phi$ and $\Omega_m$ for this critical point in
  {}Fig.~\ref{fig:evo_km1_new}. We see from the result shown in that figure 
that $\omega_{\rm tot}$ steadily tends to -1 for any initial condition, 
with scalar field density dominating over 
matter energy density ($\Omega_\phi\sim1$). Both figures~\ref{fig:numeric_evo_km1_a015}
and \ref{fig:evo_km1_new} show that the
  critical point $M_0$ is stable and yields an accelerating solution
  with $\omega_{\rm tot}\sim-1$. 

\subsection{Case III: $ n =3,\ k = 0, \ q = 0, \alpha = 0.001$}

Here, we consider the model with a constant dissipation,
  $\Upsilon_m=$ constant, which is obtained by setting $k=q=0$. Note
  that the $x^2/\Omega_m$ term, which leads to the discontinuity in
  Eq.~\eqref{u_prime} and Eq.~\eqref{v_prime}, comes with the factor
  $k$. Thus, by setting $k=0$, we no longer face the discontinuity in
  the autonomous equations and
  Eqs.~(\ref{new_x_prime})--(\ref{v_prime}) yield the critical points
  for this case. We found no critical point for
  $\alpha=0.015$. However, after lowering the value to $\alpha$ to
  0.001, we found two critical points, both of them indicating
  accelerating solutions. While $M_1$, with $\omega_{\rm tot}=-0.6$,
  is a saddle point, $M_0$ (with $\omega_{\rm tot}=-0.66$) turns out
  to be stable. We did not find any stable accelerating point with
  $\omega_{\rm tot}\sim-1$ for this case. 

\subsection{Case IV: $  n =3,\ k = 3, \ q = 0, \alpha = 0.015$}

\begin{center}
\begin{figure}[!htb]
	\includegraphics[width=10cm]{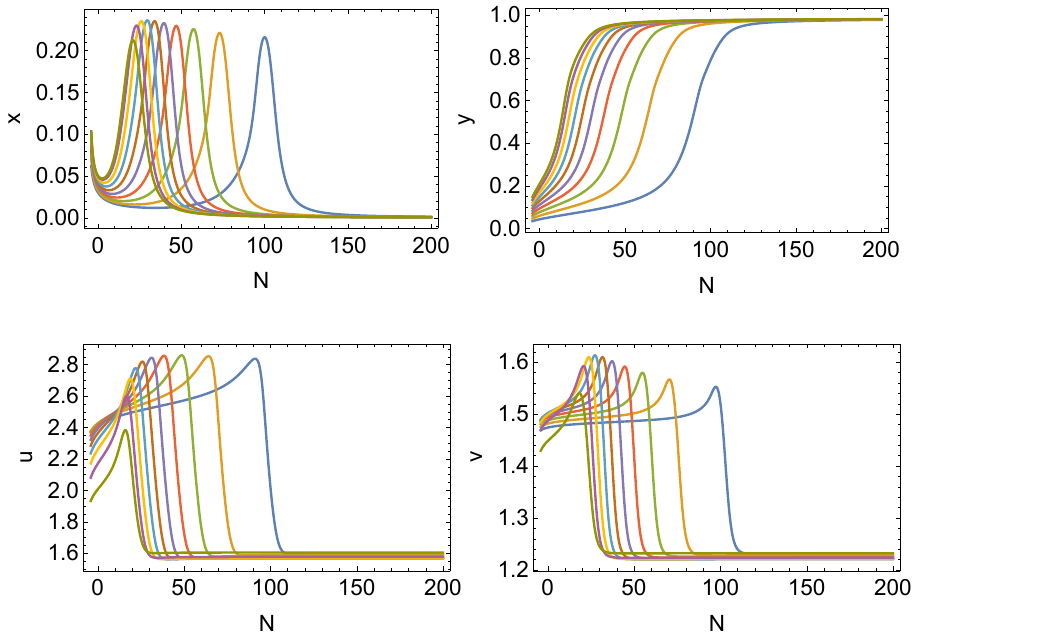}
	\caption{Numerical evolution of the autonomous equations for
          the critical point $M_0$  with \(k=3, q=0, n=3, \alpha =
          0.015\). The dynamical parameters $x$, $y$, $u$, and $v$ 
have been evolved numerically with ten different initial conditions. 
The ten different colored lines in each of these panels represent the 
evolution of these parameters with these varied initial conditions. 
The initial values of each of these parameters can be read from each 
of the plots at $N=0$.  }
	\label{fig:stab_k3_q0}
\end{figure}
\end{center}

\begin{figure}[!htb]
	\centering \includegraphics[width=7.5cm]{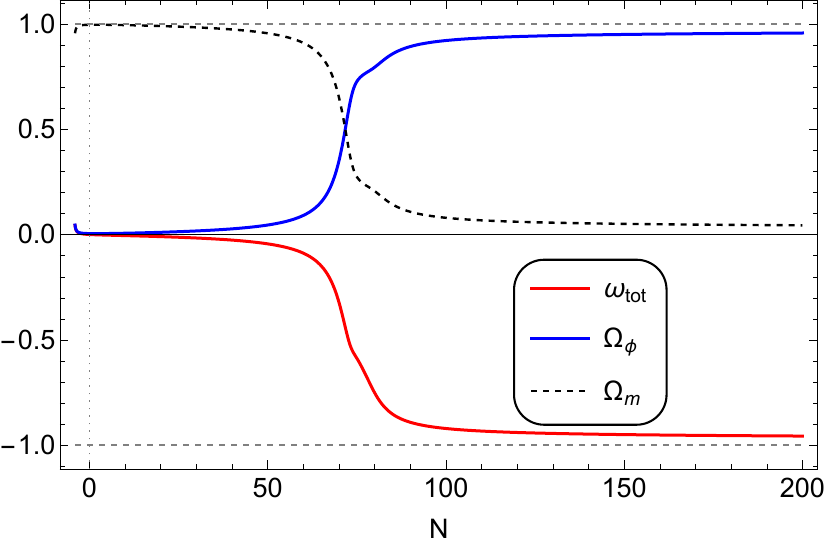}
	\caption{Evolution of cosmological parameters \( (\omega_{\rm
            tot},\ \Omega_{\phi}, \  \Omega_{m})\) corresponds to the
          critical point $M_0$ with $ k =3, \ q = 0, \ n =3,\ \alpha =
          0.015$.   }
	\label{fig:evo_k3_q0}
\end{figure}

{}Finally, we consider the model with $k=3$ and $q=0$, which yields a
  dissipation coefficient like $\Upsilon_m\propto \rho_m^{3/4}$. We
  found three critical points for this case as shown in 
Table~\ref{tab:redefined_critical_pts}. The critical point $M_{1}$ shows
  accelerating characteristics, while $M_{2}$ produces non-accelerating
  behavior at which matter energy density dominates. Both $M_1$ and
  $M_2$ turn out to be saddle points. The critical point $M_{0}$ is
  studied numerically, like in the first two models. The stability of
  the system has been evaluated numerically in
  {}Fig.~\ref{fig:stab_k3_q0}. We see that the parameters steadily 
converges to the values $(x,y,u,v) \simeq (0, 1, 1.6, 1.2)$. We plot 
the evolution of the  cosmological parameters $\omega_{\rm tot}$, $\Omega_\phi$ and
  $\Omega_m$ for this critical point  in {}Fig.~\ref{fig:evo_k3_q0},
  which shows that the model can produce a stable accelerating
  solution with $\omega_{\rm tot} \sim -1$. 

\section{Discussion and Conclusion}
\label{conclusion}

In this paper, we have studied a phenomenological model for
quintessential inflation that is motivated from warm inflation.  At
early times, the quintessential scalar inflaton field decays into
radiation during warm inflation, while at late times it is allowed to
decay into matter, thus realizing a model of dissipative interaction
in the dark sector at late times. The construction also makes use of a
generalized exponential potential able to realize both phases of
accelerated expansion, at early- and late-times.  The full dynamical
system was analyzed, with a focus  on the behavior of the dynamical
system at late times. The analysis was exemplified by both analytical
and numerical results and for different illustrative values of
parameters.  The analysis performed here extents and generalizes the
results  originally obtained in Ref.~\cite{Lima:2019yyv}, where a
version of this model was first proposed.  The results obtained
demonstrate the viability of the model as a quintessential inflation
model and in which stable solutions can be obtained. In addition, we
have also analyzed the stability of the slow-roll solutions at both
early- and late-times, which allowed us to put some constraints in the
model parameters.

\appendix

\section{Slow-roll analysis of the dynamical system}
\label{stability-slow-roll}

In this section, we shall consider the stability of the slow-roll
approximated dynamical system of the Warm Quintessential Dark Energy
Model following~\cite{Moss:2008yb}. {}From the set of equations
(\ref{eqphinew}), (\ref{eqrhoRnew}) and (\ref{eqrhom}), we see that
there are three dynamical quantities $\phi$, $\rho_r$ and $\rho_m$. We
express the radiation energy density $\rho_r$ in terms of the
entropy density $s$ as $\rho_r=(3/4)sT$ and, thus, the above set of
equations become
\begin{eqnarray}
&&\ddot\phi +3H(1+Q_r+Q_m)\dot\phi+V,_{\phi}=0,\nonumber\\ &&T\dot
  s+3HTs=3HQ_r\dot\phi^2,\nonumber\\ &&\dot\rho_m+3H\rho_m=3HQ_{m}\dot\phi^2.
\label{bckg-eom-1}
\end{eqnarray}
Under the slow-roll conditions, this set of background equations then
reduce to 
\begin{eqnarray}
&&
  3H(1+Q_r+Q_m)\dot\phi+V,_{\phi}=0,\nonumber\\ &&Ts=Q_r\dot\phi^2,\nonumber\\ &&\rho_m=Q_m\dot\phi^2.
\label{eq-slow-roll}
\end{eqnarray}
The leading order slow-roll parameters in this model
are~\cite{Moss:2008yb}
\begin{eqnarray}
\epsilon&=&\frac{1}{16\pi
  G}\left(\frac{V,_\phi}{V}\right)^2,\nonumber\\ \eta&=&\frac{1}{8\pi
  G}\frac{V,_{\phi\phi}}{V},\nonumber\\ \kappa&=&\frac{1}{8\pi
  G}\frac{V,_{\phi}}{\phi V},\nonumber\\ \beta&=&\frac{1}{8\pi
  G}\frac{V,_{\phi}\Gamma_r,_{\phi}}{V\Gamma_{r}}=p\kappa,
\nonumber\\ \gamma&=&\frac{1}{8\pi
  G}\frac{V,_\phi\Gamma_m,_\phi}{V\Gamma_m}=q\kappa,\nonumber\\ \delta&=&\frac{TV,_{\phi
    T}}{V,_{\phi}}.
\end{eqnarray}
Here, we have defined an extra slow-roll parameter $\gamma$ in
connection with the dissipation to the matter energy density, which is
in general not present in standard WI models. 

We find it convenient to change the independent variable from cosmic
time $t$ to the inflaton field $\phi$ as a clock in the equations of motion. We
also define $u\equiv \dot\phi$ and, thus, $\frac{d}{dt}=u\frac{d}{d\phi}$. Note that
this variable $u$ is different from the dynamical variable $u$ we
defined previously in Eq.~(\ref{u-and-v}). We also redefine
$\rho_m\equiv w$. Then, the set of equations given in
Eq.~(\ref{bckg-eom-1}) can be written as 
\begin{eqnarray}
&&u'=-3H-\Gamma_r-\Gamma_m-V,_\phi u^{-1}\equiv
  f(u,s,w),\nonumber\\ &&s'=-3Hsu^{-1}+T^{-1}\Gamma_r u\equiv
  g(u,s,w),\nonumber\\ &&w'=-3Hwu^{-1}+\Gamma_u u\equiv h(u,s,w),
\end{eqnarray}
where prime denotes derivative w.r.t. $\phi$.  Therefore, the
background set of equations can be compactly written as 
\begin{eqnarray}
x'=F(x),
\end{eqnarray}
where 
\begin{eqnarray}
x\equiv\left(
\begin{array}{c}
u\\ s\\ w
\end{array}
\right).
\end{eqnarray}

We take a background $\bar{x}$,  which satisfies the slow-roll
equations, Eq.~(\ref{eq-slow-roll}). Then, the linearized
perturbations satisfy the equations 
\begin{eqnarray}
\delta x'=M(\bar{x})\delta x-\bar{x},
\end{eqnarray} 
where the $M$ matrix is defined as 
\begin{eqnarray}
M=\left.\frac{\partial(f,g,h)}{\partial(u,s,w)}\right|_{u=\bar{u},s=\bar{s},w=\bar{w}}.
\end{eqnarray}
We find the matrix elements as 
\begin{eqnarray}
&&\frac{\partial f}{\partial u} =
  \frac{H}{u}\left[-3(1+Q_r+Q_m)-\frac{\epsilon}{(1+Q_r+Q_m)^2}\right]\equiv
  \mathcal{A},\nonumber\\ 
&&\frac{\partial f}{\partial s}=\frac{H}{s}\left[-cQ_r-\frac{Q_r\epsilon}{(1+Q_r+Q_m)^2}
\right.
\nonumber \\
&& \left. \;\;\;\;\;\;\;\; +\delta(1+Q_r+Q_m)\frac{}{}\right]\equiv
  \mathcal{B}, \nonumber\\ 
&& \frac{\partial f}{\partial
    w}=\frac{H}{w}\left[-\frac{3k}{4}Q_m-\frac{Q_m\epsilon}{(1+Q_r+Q_m)^2}\right]\equiv
  \mathcal{E},\nonumber\\ &&\frac{\partial g}{\partial
    u}=\frac{Hs}{u^2}\left[6-\frac{\epsilon}{(1+Q_r+Q_m)^2}\right]\equiv
          {\mathcal C},\nonumber\\ &&\frac{\partial g}{\partial
            s}=\frac{H}{u}\left[c-4-\frac{Q_r\epsilon}{(1+Q_r+Q_m)^2}\right]\equiv
          {\mathcal D},\nonumber\\ && \frac{\partial g}{\partial
            w}=\frac{Hs}{uw}\left[-\frac{Q_m\epsilon}{(1+Q_r+Q_m)^2}\right]\equiv
          {\mathcal F},\nonumber\\ &&\frac{\partial h}{\partial
            u}=\frac{Hw}{u^2}\left[6-\frac{\epsilon}{(1+Q_r+Q_m)^2}\right]\equiv
          {\mathcal G},\nonumber\\ &&\frac{\partial h}{\partial
            s}=\frac{Hw}{su}\left[-\frac{Q_r\epsilon}{(1+Q_r+Q_m)^2}\right]\equiv
          {\mathcal H},\nonumber\\ &&\frac{\partial h}{\partial
            w}=\frac{H}{u}\left[-3+\frac{3k}{4}-\frac{Q_m\epsilon}{(1+Q_r+Q_m)^2}\right]\equiv
          {\mathcal I}.
\label{matrix-elements}
\end{eqnarray}
The matrix $M$ can then be read as
\begin{eqnarray}
M=\left(
\begin{array}{ccccccc}
{\mathcal A}&&\mathcal{B}&&{\mathcal E}\\ {\mathcal C}&&{\mathcal
  D}&&{\mathcal F}\\ {\mathcal G}&&{\mathcal H}&&{\mathcal I}
\end{array}
\right).
\end{eqnarray}
The sufficient condition for stability of this slow-roll approximated
system is that the $M$ matrix varies slowly, which is justified by
having all the three eigenvalues of the diagonalized matrix to be
negative. If all the three eigenvalues of the diagonalized matrix are negative, then 
both $\det(M)$ and
${\rm tr}(M)$ should be negative as well. We find, at leading order (ignoring slow-roll
parameters),
\begin{eqnarray}
{\rm
  det}(M)&=&\frac{9}{4}\left((c-4)(4-k)+(c-4)(4+k)Q_m\right.\nonumber\\ &&\left.+(c+4)(k-4)Q_r\right),\nonumber\\ {\rm
  tr}(M)&=&-7+c+\frac{3k}{4}-3(1+Q_m+Q_r).
\end{eqnarray}
Thus, to have $\det (M)$ negative, we find the conditions $-4<c<4$ and
$-4<k<4$. These conditions also make ${\rm tr}(M)$ negative. 
We can see it explicitly that these conditions yield three negative eigenvalues of the matrix $M$ in three different physical situations: 
\begin{enumerate}
\item {\it Strong dissipative inflationary regime ($Q_r\gg1$ and $Q_m\ll1$)}: During slow-roll, with these limits, we find three eigenvalues of the matrix $M$ as $\lambda_1=(3/4)(-4+k)$, $\lambda_2=2c-3Q_r$ and $\lambda_3=-4-c$. We note that the three eigenvalues can be simultaneously negative only if $-4<c<4$ and $-4<k<4$.

\item {\it Weak dissipative inflationary regime ($Q_r\ll1$ and $Q_m\ll1$)}: In this case, we find the three eigenvalues as $\lambda_1=-3$, $\lambda_2=-4+c$ and $\lambda_3=(3/4)(-4+k)$. Here also, we note that the conditions to get all the three eigenvalues negative are $-4<c<4$ and
$-4<k<4$.

\item {\it Quintessence driven Dark Energy dominated regime ($Q_r\ll1$ and $Q_m\gg1$)}: Here the three eigenvalues turn out to be $\lambda_1=-4+c$, $\lambda_2=(3/2)(k-2Q_m)$ and $\lambda_3=-(3/4)(4+k)$. Like in the previous two cases, in this case too, the conditions to get all the three eigenvalues negative are $-4<c<4$ and
$-4<k<4$.
\end{enumerate}
Therefore, we
see that for the system to be stabilized, the form of the dissipative
coefficients $\Upsilon_r$ and $\Upsilon_m$ must involve the powers $c$ and $k$ 
satisfying the conditions $-4<c<4$ and
$-4<k<4$.

\begin{acknowledgments}

R.O.R. acknowledges financial support  by research grants from
Conselho Nacional de Desenvolvimento Cient\'{\i}fico e Tecnol\'ogico
(CNPq), Grant No. 307286/2021-5, and from Funda\c{c}\~ao Carlos Chagas
Filho de Amparo \`a Pesquisa do Estado do Rio de Janeiro (FAPERJ),
Grant No. E-26/201.150/2021.  R.S. was supported by a scholarship from
FAPERJ.

\end{acknowledgments}



\end{document}